\documentclass{elsart}
\usepackage{epsfig,psfrag}

\newcommand{\I}{\mathrm{i}}

\begin{document}

\begin{frontmatter}

\title{Electron transfer rates for asymmetric reactions}

\author[lothar]{L.~M{\"u}hlbacher}
\ead{lothar.muehlbacher@physik.uni-freiburg.de}
\author[reinhold]{and R.~Egger}
\address[lothar]{Physikalisches Institut,
 Albert-Ludwigs-Universit{\"a}t,
 D-79104 Freiburg}
\address[reinhold]{Institut f{\"u}r Theoretische Physik,
 Heinrich-Heine-Universit{\"a}t,
 D-40225 D{\"u}sseldorf}

\begin{abstract}
We use a numerically exact real-time path integral Monte Carlo scheme to
compute electron transfer dynamics between two redox sites within a spin-boson
approach. The case of asymmetric reactions is studied in detail in the least
understood crossover region between nonadiabatic and adiabatic electron
transfer. At intermediate-to-high temperature, we find good agreement with
standard Marcus theory, provided dynamical recrossing effects are captured.
The agreement with our data is practically perfect when temperature
renormalization is allowed. At low temperature we find peculiar electron
transfer kinetics in strongly asymmetric systems, characterized
by rapid transient dynamics and backflow to the donor.
\end{abstract}

\end{frontmatter}

\section{Introduction}

This paper presents results of computer simulations for simple asymmetric
electron transfer (ET) reactions in polar solvents. Such reactions are
encountered in a large variety of different systems, e.g.~charge transfer in
semiconductors, chemical reactions, or the primary step in bacterial
photosynthesis \cite{marcus85,zusman,kuznetsov,chandler,weiss,tributsch}. Our
theoretical study of such processes is based on the spin-boson (SB) model
(dissipative two-level system) \cite{leggett}. This model provides an accurate
description of many ET reactions involving condensed-phase environments
\cite{chandler}, if a suitable spectral density $J(\omega)$ can be determined
for these ``bath'' modes. For a detailed discussion of restrictions for the SB
description of ET reactions, including an (almost) exhaustive list of
references to work done in that context, we refer to our previous work
\cite{latest_one} for symmetric ET reactions. In the present paper, primary
focus is laid on the asymmetric case. Within the SB model, the localized sites
representing the donor and the acceptor are described in terms of an effective,
quantum mechanical ``spin'' variable. The two-level system (TLS) is
characterized by a tunnel splitting $\hbar \Delta$, which is twice the usual
electronic coupling between the redox sites, and by the asymmetry
$\hbar\epsilon$ giving the energy difference between the two localized states.
The frequency-resolved coupling strength of the electron to the surrounding
solvent modes is then described by the spectral density, where one mimics the
environment by an infinite set of effective harmonic oscillators. The spectral
density for a given ET reaction can in principle be computed from classical
molecular dynamics simulations \cite{chandler}, but in many cases it turns out
to be appropriate to consider the special class of Ohmic spectral densities
\cite{weiss,leggett},
\begin{equation} \label{ohmic}
J(\omega) = 2\pi\alpha\omega\, \e^{-\omega/\omega_c} ,
\end{equation}
with dimensionless damping strength $\alpha$ and a cutoff frequency $\omega_c$.
On general grounds, the linear low-frequency behavior of $J(\omega)$ is
expected in basically all condensed-phase ET reactions \cite{chandler,weiss},
and the frequency $\omega_c$ then corresponds to some dominant bath mode. To
make contact with common notation, we shall use the reorganization energy
$\hbar \Lambda$ \cite{marcus85} instead of $\alpha$, where $\Lambda=2\alpha
\omega_c$. While our method below is able to treat arbitrary spectral
densities, for clarity we will only present data using Eq.~(\ref{ohmic}). Note
that for common ET reactions, one has large damping parameters $\alpha$, and
therefore one is always in the incoherent regime with respect to the electronic
degree of freedom. Coherence is nevertheless possible with respect to the bath
degrees of freedom, in particular when fast bath modes are absent or particular
initial preparations are selected \cite{lucke,lucke2}. We mention in passing
that due to the strong damping, weak-coupling Redfield-type approaches
\cite{friesner} are not suitable for a description of condensed-phase ET
processes.

For very large $\omega_c/\Delta$ and sufficiently low temperatures, the scaling
limit of the SB model is reached, where powerful analytical methods are
available \cite{weiss,leggett}. Unfortunately, the scaling regime is of little
relevance to ET reactions, but in certain parameter regions, progress is
nevertheless possible. A famous and very successful description of ET reactions
concerns the high-temperature limit, where the bath behaves classically,
leading to Marcus theory \cite{marcus85}. Indeed, it can be shown that the
classical limit of the SB model directly gives Marcus theory
\cite{garg85,song}. Another example is given by the golden rule formula for the
nonadiabatic ET rate \cite{levich,GR2}, which is accurate for arbitrary
temperature as long as $\Delta/\omega_c\ll 1$. In the opposite adiabatic limit
of a very slow (classical) bath, analytical progress is also possible
\cite{carmeli}. A particular focus of our study lies on the crossover regime
between adiabatic and nonadiabatic ET, which represents the most difficult and
least understood regime from a theory point of view, in particular at low
temperatures.

In our previous work \cite{latest_one}, exactly this regime was studied for
symmetric systems. Applying the path-integral Monte Carlo (PIMC) scheme of
Ref.~\cite{latest_one} to $\epsilon\neq 0$, we are able to directly monitor the
ET dynamics without any approximation. This should be contrasted to alternative
numerical routes to this problem. For instance, in mixed
quantum/(semi)classical simulations for the SB model
\cite{stock3,miller,golosov2,thoss,lucke3}, it is hard to justify some of the
approximations, and hence the accuracy of the results has to be checked case by
case. Similar arguments apply to basis set calculations \cite{wang}, and to
memory-equation approaches \cite{nancy,winterstetter}. Other methods are known
to be accurate, but only in either the scaling regime or the weak-coupling
regime \cite{stockburger1,stockburger2,costi1,schoeller}. We conclude that the
PIMC method represents an excellent computer simulation method for treating the
crossover regime mentioned above. PIMC is free from approximations, but has to
deal with the well known dynamical sign problem, because a calculation of ET
rates requires dynamical information incorporating interfering real-time
trajectories. Some workers have attempted to use analytical continuation of
imaginary-time PIMC data \cite{sudip,voelker,bailey}, but this process is
mathematically ill-defined and troublesome to carry out. Here we instead
directly proceed in real time, where reliable error estimates can be obtained
and true nonequilibrium preparations can be investigated \cite{latest_one,acp}.
The outline of the paper is as follows. In Sec.~\ref{sec2}, we summarize the
model and the simulation technique. Results for asymmetric ET reactions are
presented in Sec.~\ref{sec3}. We close with a concluding discussion in
Sec.~\ref{sec4}.

\section{Spin-boson description of ET reactions} 
\label{sec2}

The spin-boson model \cite{weiss,leggett} is defined by a coupled system-bath
Hamiltonian $H=H_0+H_I+H_B$. The ``spin'' corresponding to the TLS is
parametrized by Pauli matrices $\sigma_i$, where the $|+\rangle$ ($|-\rangle$)
eigenstate of $\sigma_z$ refers to the donor (acceptor). In the absence of the
solvent, this leads to
\begin{equation}\label{h0}
H_0 = -{\hbar\Delta\over2}\sigma_x + {\hbar\epsilon\over2}\sigma_z.
\end{equation}
The environmental modes are modeled by an infinite collection of harmonic
oscillators, $H_B$, which bilinearly couple to the position of the electron
($H_I$). The solvent-TLS coupling is completely encoded in a spectral density
$J(\omega)$, which effectively becomes a continuous function of $\omega$ for
condensed-phase environments, and determines all bath correlation functions
that are relevant for ET dynamics. Here we take the Ohmic form (\ref{ohmic}) as
a prototype spectral density. The bath autocorrelation function for complex
time $z=t-\I\tau$ is (with $\beta=1/k_{\rm B} T$)
\begin{equation} \label{lz}
L(z) = \int_0^\infty \frac{\d\omega}{\pi} J(\omega)
{\cosh[\omega(\hbar\beta/2-\I z)] \over \sinh(\hbar\beta\omega/2)}.
\end{equation}
In the classical limit, the reorganization energy $\hbar \Lambda$, which is an
integral quantity describing the overall coupling strength, is the only
important bath quantity. Details of the frequency dependence of $J(\omega)$
become crucial only at low temperatures.

Here we study two dynamical properties characterizing ET kinetics. First,
\begin{equation} \label{P(t)}
P(t) = \langle\sigma_z(t)\rangle = \langle \e^{\I Ht/\hbar} \sigma_z
\e^{-\I Ht/\hbar} \rangle
\end{equation}
gives the difference in occupation probabilities of the donor and the acceptor
state, with the electron initially fixed on the donor. This quantity then
directly probes ET dynamics after a nonequilibrium initial preparation,
e.g.~following photoexcitation of an electron into the donor state. Second, the
correlation function
\begin{equation} \label{corr}
C(t) = \langle\sigma_z(0)\sigma_z(t)\rangle_\beta
= Z^{-1}{\rm tr}\left\{
 \e^{-\beta H}\sigma_z \e^{\I Ht/\hbar}\sigma_z \e^{-\I Ht/\hbar}
 \right\}
\end{equation}
with $Z = {\rm tr}\{\exp(-\beta H)\}$ probes the dynamics under an equilibrium
preparation. At high temperatures, preparation effects are not expected to
dramatically affect the ET dynamics, and therefore a well-defined thermal rate
$k_{\rm th}$ should exist \cite{VCM,hanggi}. This rate then supposedly governs
the exponential relaxation of both $P(t)$ and $C(t)$ on a timescale of the
order of $\tau_{\rm relax} \approx k_{\rm th}^{-1}$. 
In principle, at low temperatures, this assertion can be violated, and
then the relevant quantity of interest depends on the experimental setup under
consideration.

In case a unique rate constant exists, it is composed of the  forward rate
$k_\mathrm{f}$
and the  backward rate $k_\mathrm{b}$, referring to directed transfers
between the donor and acceptor states according to
\[
dP_+ = (-k_f P_+ + k_b P_-)dt\;,\quad dP_- = (-k_b P_- + k_f P_+)dt
\]
for times beyond some transient timescale,
 $t > \tau_{\rm trans}\approx 1/\omega_{\mathrm{c}}$. 
Under such a rate process,  ET dynamics obeys
\begin{equation} \label{dynamics}
P(t>\tau_{\rm trans}) = P_{\rm trans}\exp(-k_{\rm th}t) + P_\infty,
\end{equation}
where $P_\infty = \langle\sigma_z\rangle_\beta$ denotes the electronic
equilibrium population, 
\[
P_{\rm trans} = [P(\tau_{\rm trans}) -
P_\infty]\exp(k_{\rm th}\tau_{\rm trans}),
\]
 and $k_\mathrm{th}$ is the total
transfer rate $k_\mathrm{f}+k_\mathrm{b}$. Accordingly, $P(t)$ follows a simple
exponential decay for times $t > \tau_{\rm trans}$.
 
Another way to obtain the thermal transfer rate
$k_{\rm th}$ is via a time-dependent equilibrium rate function
\cite{VCM,hanggi}. Assuming that $k_\mathrm{f}$ and $k_\mathrm{b}$ are
connected via a standard detailed balance relation,
\begin{equation} \label{detailed balance}
k_\mathrm{b} = k_\mathrm{f} \exp(-\hbar\beta\epsilon) \;,
\end{equation}
for the SB model this function reads
\cite{latest_one}
\begin{equation} \label{timerate}
k(t) = {1 + \cosh(\hbar\beta\epsilon) \over \hbar\beta}
{\rm Im} C(t) ,
\end{equation}
where we exploited that Eq.~(\ref{detailed balance}) corresponds to the
assumption of a Boltzmann distribution for the equilibrium occupation
probabilities, $P_\infty = -\tanh(\hbar \beta\epsilon/2)$.  This is
expected to always hold 
under the strong damping conditions characteristic for ET reactions.
Indeed, our
numerical results below verify this assumption to high precision.

\begin{figure}
\centering
\hspace*{0cm}\includegraphics[height=70mm,draft=false]{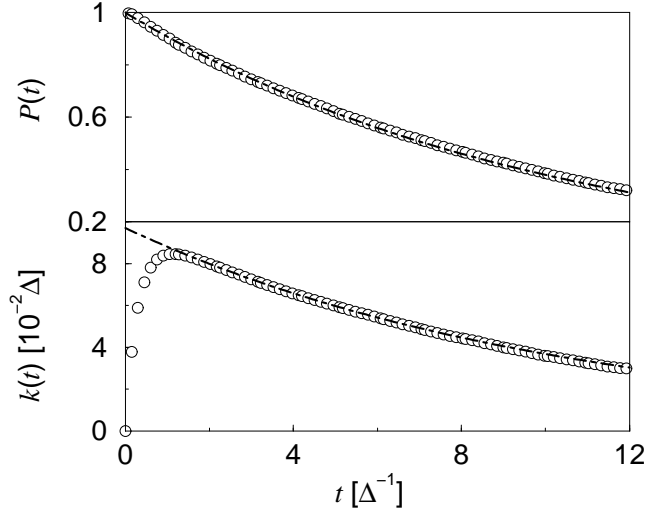}
\caption[]{\label{fig_1} ET dynamics for $\Lambda/\Delta =15$, $\epsilon=0$,
$\Delta/\omega_\mathrm{c}=0.067$, and $T=\hbar\Delta/k_{\rm B}$.  Shown are
PIMC data for $P(t)$ (top) and $k(t)$ (bottom).  Dashed curves are
$\exp(-k_{\rm th}t)$ (top) and $k_{\rm th}\exp(-k_{\rm th}t)$ (bottom), with
$k_{\rm th} = 0.097\Delta$. Error bars (one standard deviation) are smaller
than symbol size. }
\end{figure}

According to the reasoning in Ref.~\cite{VCM}, $k_{\rm th}$ is then obtained as
the plateau value of $k(t)$ if such a plateau exists. However, a careful
re-examination of the argument in Ref.~\cite{VCM}, see also
Ref.~\cite{chandbook}, reveals that this procedure is just the limiting case of
a more general situation, where $k(t)$ decays exponentially according to
\begin{equation}\label{ratefit}
k(t>\tau_{\rm trans})=k_{\rm th}e^{-k_{\rm th}t}.
\end{equation}
Therefore, even if Eq.~(\ref{timerate}) exhibits no clear plateau, a reaction
rate can still be extracted if the ``exponential fit'' to Eq.~(\ref{ratefit})
is possible. A concrete example is shown in Fig.~\ref{fig_1}, where the PIMC
data for $P(t)$ and $k(t)$ are both consistent with the same thermal rate
constant $k_{\rm th}=0.097 \Delta$. The exponential fit to Eq.~(\ref{ratefit})
is generally more accurate than extracting $k_{\rm th}$ directly from $P(t)$,
where transient decays for $t<\tau_{\rm trans}$ complicate the analysis. The
latter approach was taken in Ref.~\cite{latest_one} whenever no plateau was
found for $k(t)$, but then the rates could have been obtained to better
precision using the exponential fitting procedure for $k(t)$. We mention in
passing that regarding Fig.~7 of Ref.~\cite{latest_one}, ET rates for $T \ge
2\hbar\Delta/k_{\rm B}$ indeed closely follow the classical Marcus rate, see
Eq.~(\ref{marcus}) below, if one takes a renormalized temperature $T' = 0.87T$
($0.78T$) for $\Lambda/\Delta=10$ and $\Delta/\omega_\mathrm{c}=1$
($\Delta/\omega_\mathrm{c}=2$). Obviously, fits to Eq.~(\ref{ratefit}) can
fail, in particular when (vibrational) coherence survives, or when $\tau_{\rm
trans}$ and $\tau_{\rm relax}$ are not sufficiently well separated. Such
problems will then indicate the invalidity of a simple rate picture in these
cases.

Before describing the PIMC method employed here, we briefly review classical
Marcus theory \cite{marcus85}. Writing the thermal transfer rate in terms of
the forward rate, $k^{\rm cl} = k^{\rm cl}_{\rm f} (1+\e^{-\hbar\epsilon/k_{\rm
B}T})$, Marcus theory predicts
\begin{equation} \label{marcus}
k^{\rm cl}_{\rm f}
=\frac{\Delta^2}{4+\pi\Delta^2/(\Lambda\omega_{\rm r})} \sqrt{{\pi \hbar \over
\Lambda k_{\rm B} T}}\, \e^{-F^\ast(\epsilon)/k_{\rm B} T} ,
\end{equation}
with the activation free energy barrier (Marcus
parabola)
\begin{equation} \label{Fstar}
F^\ast(\epsilon) = \hbar(\epsilon - \Lambda)^2/4\Lambda,
\end{equation}
and a solvent frequency scale $\omega_{\rm r}$. The $\omega_{\rm r}$ dependent
prefactor in Eq.~(\ref{marcus}) was introduced by Zusman \cite{zusman} to
account for recrossing events. The scale $\omega_{\rm r}$ seems to follow a
power law at low temperatures \cite{latest_one}, but in the true classical
(high-temperature) regime, it is given by $\omega_{\rm r}= \omega_\mathrm{c}/2$
\cite{latest_one,garg85}. Marcus theory (supplemented by dynamical recrossing
effects) yields a classical rate constant covering the full crossover from
nonadiabatic to adiabatic ET. One of its most striking features is a
non-monotonic behavior of the forward rate as a function of asymmetry
$\epsilon$. According to Eq.~(\ref{Fstar}), $k^{\rm cl}_{\rm f}$ exhibits its
single maximum in the activationless case $\epsilon = \Lambda$, but decreases
again in the inverted regime $\epsilon > \Lambda$. While experiments
qualitatively support this picture \cite{miller_calcaterra}, quantum effects
(nuclear tunneling) not captured by Marcus theory should play a major role in
the inverted regime \cite{marcus85}. Such effects are of course included by
the PIMC simulations below.

Next we briefly describe our computational technique used to calculate ET rate
constants (if they exist) under a spin-boson description. A detailed
exposition of our method can be found in Ref.~\cite{latest_one}, and here we
only sketch it to keep the paper self-consistent. After tracing out the
Gaussian bath degrees of freedom in Eqs.~(\ref{P(t)}) and (\ref{corr}), one
arrives at path-integral expressions for the dynamical quantities of interest,
where only the TLS degree of freedom is kept explicitly. For instance,
\begin{equation} \label{corr2}
C(t) = Z^{-1} \int\!{\mathcal D}\sigma\; \sigma(0)\sigma(t)
 \exp\left\{ {\I\over\hbar}S_0[\sigma] - \Phi[\sigma] \right\} ,
\end{equation}
where the path integration runs over paths $\sigma(z)$
 for the discrete spin
variable $\sigma=\pm 1$ corresponding to the eigenvalues of $\sigma_z$, with
the complex time $z$ following the Kadanoff-Baym contour $\gamma = \{z: 0
\rightarrow t \rightarrow 0 \rightarrow -\I\hbar\beta\}$. 
$S_0[\sigma]$ denotes the free TLS action, while the effective action
due to the 
traced-out bath is captured by the {\sl influence functional} \cite{weiss}
\begin{equation} \label{inflfunct}
\Phi[\sigma] = {1\over4} \int_\gamma \d z \int_{z'<z} \d z' \sigma(z) L(z-z')
\sigma(z') ,
\end{equation}
with integrations ordered along the contour $\gamma$. For a numerical
evaluation, time is discretized, and, moreover, the real-time spins
$\sigma(t')$ and $\sigma'(t')$ on the forward and backward branch of the
Kadanoff-Baym contour $\gamma$ are combined to form quantum fluctuations
($\sigma-\sigma'$) and quasiclassical variables ($\sigma+\sigma'$)
which are sampled stochastically in the course of the
PIMC simulation. Technical details about our implementation of this program
can be found in Ref.~\cite{latest_one}.

While similar expressions can be derived for the occupation probability, $P(t)$
can also be obtained from the same MC trajectory as $C(t)$ (and therefore
$k(t)$) by including a correction factor that accommodates the changes in the
corresponding MC weight \cite{latest_one}. In principle, this allows to
simultaneously compute $k(t)$ and $P(t)$ without significant increase in
computing time. However, for a strongly biased system, $\epsilon/\Lambda \gg
1$, this approach has a rather poor performance since the respective MC weights
refer to almost orthogonal initial electronic states. It would then be
necessary to run separate simulations to extract $P(t)$ and $k(t)$. More
serious complications arise from the fact that $k(t)$ in Eq.~(\ref{timerate})
is composed of two competing contributions, namely
$[1+\cosh(\hbar\beta\epsilon)]$ and ${\rm Im}C(t)/\hbar\beta$. Since the former
increases exponentially with $|\epsilon|$, the latter should be very tiny in
order to give a finite result. Unfortunately, our simulations show that the
stochastic PIMC error in ${\rm Im}C(t)$ is rather insensitive to a change of
parameters, as long as the time scale of the simulation is kept fixed. Hence
relative errors become arbitrarily large as ${\rm Im}C(t)\to 0$, which renders
data for $k_{\rm th}$ numerically unstable. Since the relevant parameter is
$|\hbar\beta\epsilon|$, extraction of $k_{\rm th}$ from $k(t)$ fails already at
rather small $|\epsilon|$ for low temperature. These numerical problems are
unrelated to the dynamical sign problem, which did not impose serious
limitations in the parameter regime investigated here. Therefore, there was no
need to employ the powerful but complicated ``multilevel blocking'' algorithm
\cite{latest_one} to relieve the sign problem.

\section{Numerical results for asymmetric ET}
\label{sec3}

\begin{figure}
\centering
\psfrag{E}{$\epsilon$}
\includegraphics[height=75mm,draft=false]{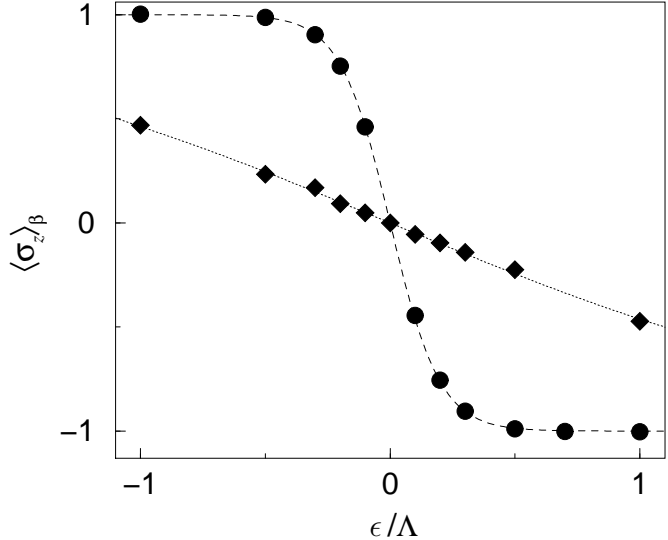}
\caption[]{\label{fig_4} Equilibrium
expectation value $\langle\sigma_z\rangle_\beta$ as a function of 
$\epsilon$ for $T=\hbar\Delta/k_{\rm B}$ (circles) and 
$T=10\hbar\Delta/k_{\rm B}$
(diamonds). The dashed and dotted lines refer to $-\tanh(\hbar\beta\epsilon/2)$.
Solvent parameters are $\Delta/\omega_\mathrm{c} = 1$ and $\Lambda =
10\Delta$. }
\end{figure}

Being mainly interested in the effect of the asymmetry $\epsilon$ on the ET
dynamics, we have calculated both the occupation probability $P(t)$ and the
time-dependent rate function $k(t)$ for various values of $\epsilon$ at a high
and a low temperature. Specifically, we consider $T=10\hbar\Delta/k_{\rm B}$
and $T=\hbar\Delta/k_{\rm B}$, and take bath parameters where Marcus theory has
been shown to be accurate in the symmetric case \cite{latest_one}, namely
$\Delta/\omega_\mathrm{c} = 1$ and $\Lambda = 10\Delta$. This parameter choice
represents the most interesting crossover region between nonadiabatic and
adiabatic ET.

For these temperatures, from closer inspection of Eq.~(\ref{marcus}), the rate
$k_{\rm th}(\epsilon)$ is expected to exhibit a single maximum at $\epsilon =
0$ for the higher temperature, but two symmetric maxima at $\epsilon=\pm
\epsilon_{\rm max}$ with $\epsilon_{\rm max} < \Lambda$ for the lower one. From
Eq.~(\ref{marcus}), it is easy to see that $\epsilon_{\rm max}$ is a solution
of the equation
\[
(1-\epsilon/\Lambda) [1+\exp(\hbar\beta\epsilon)]=2 .
\]
Furthermore, due to Eq.~(\ref{detailed balance}), the thermal transfer rate is
a symmetric function of $\epsilon$. Quantum corrections to the classical rate
(\ref{marcus}) are then of primary interest to us,
in particular in the inverted regime, $|\epsilon| >\Lambda$,
for both signs of the asymmetry.
Before turning to the thermal
transfer rate, we show the equilibrium expectation value of $\sigma_z$ as a
function of $\epsilon$, see Fig.~\ref{fig_4}. For the parameters investigated,
we find to high precision $\langle\sigma_z\rangle_\beta =
-\tanh(\hbar\beta\epsilon/2)$, as has already been mentioned above.

\begin{figure}
\centering
\psfrag{E}{$\epsilon$}
\includegraphics[height=75mm,draft=false]{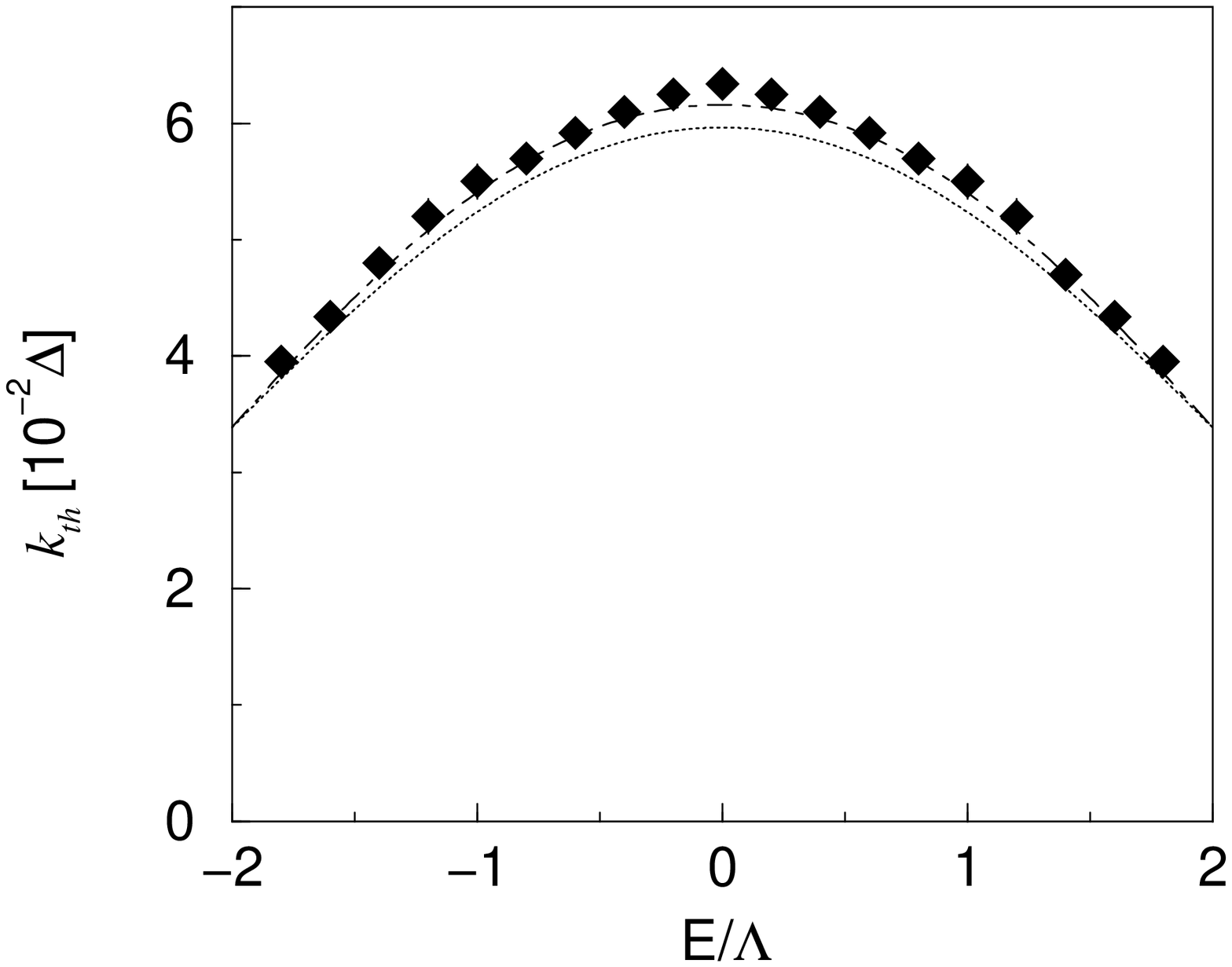}\\
\includegraphics[height=75mm,draft=false]{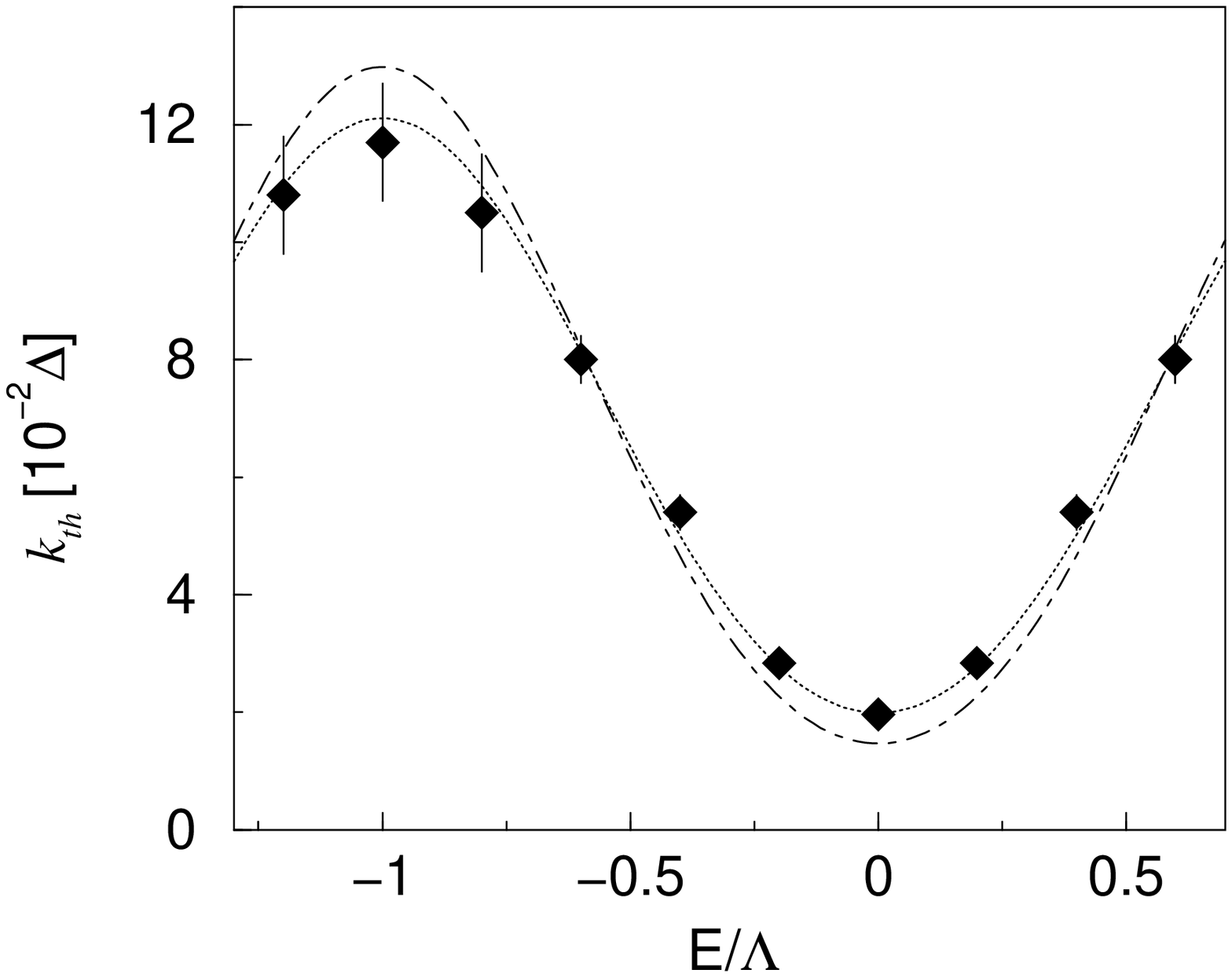}
\caption[]{\label{fig_5} $k_{\rm th}(\epsilon)$ for $T=10 \hbar\Delta/k_{\rm
 B}$ (top) and $T=\hbar\Delta/k_{\rm B}$ (bottom). The dotted (dashed) curve
 refers to Eq.~(\ref{marcus}) without (with) a renormalization of the
 temperature, $T'=0.87T$. The larger error bars for $T=\hbar\Delta/k_{\rm B}$
 and $\epsilon/\Lambda \le -0.8$ are caused by rate extraction from $P(t)$
 rather than $k(t)$. }
\end{figure}

\begin{figure}
\centering
\hspace*{-1.5cm}\includegraphics[height=75mm,draft=false]{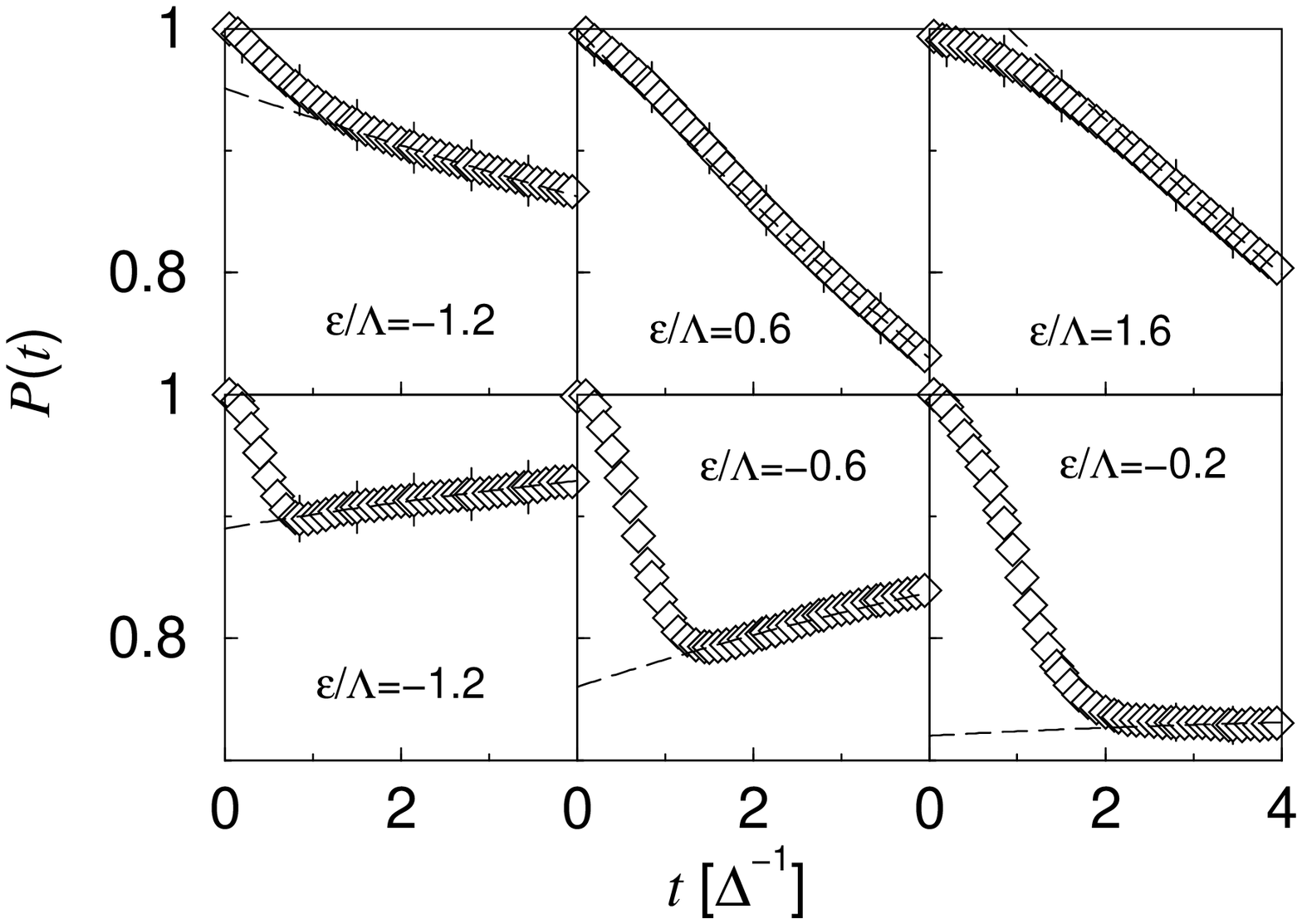}
\caption[]{\label{fig_6} $P(t)$ for $T=10\hbar\Delta/k_{\rm B}$ (top) and
$T=\hbar\Delta/k_{\rm B}$ (bottom). The dot-dashed curves denote  fits
to Eq.~(\ref{dynamics}) with 
$k_{\rm th}$ as in Fig.~\ref{fig_5} and
(top left to top right) $P_{\rm trans} = 0.41$, $1.29$, $1.72$, and (bottom
left to bottom right) $-0.11, -0.24, -0.04$, respectively. For the sake
of clarity, only  part of the data is shown. }
\end{figure}

Starting with the high temperature $T=10\hbar\Delta/k_{\rm B}$, we then
calculated the ET dynamics for $|\epsilon|/\Lambda \le 1.8$ while keeping all
other parameters fixed. Then the difficulties mentioned above did not pose
serious limitations. The corresponding data for $k_{\rm th}$, which always
follow from exponential fits to $k(t)$, are displayed in
Fig.~\ref{fig_5}. Similar to the crossover regime for the symmetric system
\cite{latest_one}, the classical rate~(\ref{marcus}) again nicely reproduces
the exact results. Our PIMC data slightly exceed the classical prediction,
especially for small $|\epsilon|$. However, these deviations can be
astonishingly well compensated for by the above-mentioned renormalization of
the temperature. While we do not have a good theoretical argument for the
validity of the temperature renormalization, we wish to stress that a
renormalization of the reorganization energy and/or the activation energy is
less apt to fit our data over the full asymmetry range, and
moreover, completely fails to capture the mentioned deviations 
in the symmetric case.
Notably, in the high-temperature regime, no significant trace of nuclear
tunneling could be observed in the inverted regime. Outside the transient
motion, ET dynamics exhibits the exponential decay (\ref{dynamics}), with
slower (faster) decay than during the transient stage for $\epsilon \le 0$
($\epsilon > 0$), see Fig.~\ref{fig_6}.

For the lower temperature $T=\hbar\Delta/k_{\rm B}$, the numerical limitations
discussed above became quite severe. Extraction of the transfer rate from
$k(t)$ could only be achieved for $|\epsilon|/\Lambda \le 0.6$. Nevertheless,
for $\epsilon < 0$, $k_{\rm th}$ could still be obtained from $P(t)$, although
with poorer accuracy since two fit parameters ($k_{\rm
th}$ and $P_{\rm trans}$) are required.  This extends the range of our data to
$-1.4 \le \epsilon/\Lambda \le 0.6$, and hence includes one of the rate
maxima. Again the classical rate yields a remarkably accurate prediction for
$k_{\rm th}$, see Fig.~\ref{fig_5}, while the temperature renormalization is
seen to be limited to the high-temperature regime \cite{latest_one}. 
One might therefore be tempted to conclude that there are no 
significant quantum effects even at low temperatures.
However, for $\epsilon/\Lambda < -0.2$, the ET dynamics shows a
qualitatively different behavior. 
Due to the strong equilibrium localization of
the electron to the donor state, the fast low-temperature
initial transient kinetics is able to carry
$P(\tau_{\rm trans})$ to a value below $P_\infty$, causing a subsequent
\emph{increase} of $P(t)$ towards $P_\infty$, see
Fig~\ref{fig_6}. Nevertheless, the subsequent time evolution is still correctly
captured by Eq.~(\ref{dynamics}), albeit the exponentially decaying excess
population with respect to its equilibrium value is now
negative.  This effect nicely illustrates that even
outside the transient kinetics, only calculating the transfer rate is 
sometimes not sufficient to fully characterize the ET dynamics.

\section{Discussion}
\label{sec4}

Using the SB model as a description for ET processes, we have calculated the
thermal transfer rate $k_{\rm th}$ for a bath setup in the crossover regime,
where Marcus theory is known to provide a valid description for the symmetric
electronic system. Focusing on the influence of the energy bias $\epsilon$
between the redox states, we investigated the electronic dynamics for two
different temperatures, $T=\hbar\Delta/k_{\rm B}$ and $T=10\hbar\Delta/k_{\rm
B}$ and various values of $\epsilon$ up to the inverted regime using real-time
PIMC simulations. We extended our previous approach to extract the thermal
transfer rate from the time-dependent function $k(t)$, see
Eq.~(\ref{timerate}), which yields a powerful tool both to obtain
$k_{\rm th}$ and to decide whether a rate description is appropriate in the
first place. In the absence of electronic or vibrational coherence, this rate
also describes the exponential decay of the electronic population $P(t)$,
provided the timescales for transient dynamics $\tau_{\rm trans}$ and thermal
relaxation $\tau_{\rm relax}$ are sufficiently well separated.

Our results for the asymmetric electronic system support the rate picture at
all values of the electronic bias investigated, which extend well into the
inverted regime. This also applies for the cases where the transient motion
shifts the electronic population $P(t)$ below its equilibrium value, i.e.~at
$\epsilon < 0$ and sufficiently low temperatures, where it is the subsequent
deviation of $P(t)$ from this equilibrium value which decays
exponentially. Furthermore, the validity of the classical rate~(\ref{marcus})
was tested as a function of the asymmetry. Deviations from
the classical rate expression indicating contributions from nuclear tunneling
were largely absent, 
suggesting that they should only play a major role at either considerably lower
temperatures than $T = \hbar\Delta/k_{\rm B}$, or for
very large asymmetry  $|\epsilon|/\Lambda$.
However, we could observe an overall enhancement of the rate which can be
nicely captured by evaluating the classical rate for a renormalized 
temperature $T' < T$.
This phenomenon was also confirmed for the symmetric
system,  seems to be confined to the high-temperature regime,
and becomes more pronounced towards the adiabatic limit.
Furthermore, at low temperatures we found a 
rapid transient dynamics, followed by a backflow to the donor.
This effect shows that a naive rate picture breaks down in the
low-temperature regime, although the timescale for relaxation is
still well predicted by the classical prediction.
These findings illustrate that despite the good accuracy of classical Marcus
theory, which is confirmed once again by our paper, 
a fully quantum-mechanical description of the bath's influence on the ET
process can be important.
For the lower temperature, our method eventually failed to produce reliable
data with increasing $|\epsilon|$, especially for $\epsilon > 0$. While there
is no obvious cure for the computation of $k(t)$, for $P(t)$ these problems
seem to mainly stem from our use of a particular PIMC scheme 
optimized for computation of
$k(t)$.  A direct calculation
of $P(t)$ will be free of these problems and could therefore allow to
significantly extend the range of asymmetries. 

This work has been supported by the Volkswagen-Stiftung and by the Deutsche
Forschungsgemeinschaft.


\begin{thebibliography}{00}

\bibitem{marcus85}
R.A. Marcus and N. Sutin, Biochim. Biophys. Acta 811 (1985) 265.

\bibitem{zusman}
L.D. Zusman, Chem. Phys. 49 (1980) 295.

\bibitem{kuznetsov}
A.M. Kuznetsov, Charge transfer in physics, chemistry and biology
(Gordon and Breach, 1995).

\bibitem{chandler}
D. Chandler, in: Liquids, Freezing and the Glass
Transition, edited by D. Levesque et al. (Elsevier
Science, North Holland, 1991).


\bibitem{weiss}
U. Weiss, Quantum Dissipative Systems, 2nd edition
 (World Scientific, Singapore, 1998).

\bibitem{tributsch}
H. Tributsch and L. Pohlmann, Science 279 (1998) 1891.

\bibitem{leggett}
A.J. Leggett, S. Chakravarty, A.T. Dorsey, M.P.A. Fisher, A. Garg, and
W. Zwerger, Rev. Mod. Phys. 59 (1987) 1.

\bibitem{latest_one}
L. M{\"u}hlbacher and R. Egger, J. Chem. Phys. 118 (2003) 179.
See also: R. Egger, L. M{\"u}hlbacher, and C.H. Mak, Phys. Rev. E
61 (2000) 5961.

\bibitem{lucke}
A. Lucke, C.H. Mak, R. Egger, J. Ankerhold, 
J. Stockburger, and H. Grabert, J. Chem. Phys. 107 (1997) 8397.

\bibitem{lucke2}
A. Lucke and J. Ankerhold, J. Chem. Phys. 115 (2001) 4696.

\bibitem{friesner}
W.Th. Pollard, A.K. Felts, and R.A. Friesner,
Adv. Chem. Phys. 93 (1996) 77.

\bibitem{garg85}
A. Garg, J.N. Onuchic, and V. Ambegaokar, 
J. Chem. Phys. 83 (1985) 4491.

\bibitem{song}
X. Song and A.A. Stuchebrukhov, J. Chem. Phys.
99 (1993) 969.

\bibitem{levich}
V.G. Levich, Adv. Electrochem. Electrochem. Eng. 4 (1965) 249.

\bibitem{GR2}
R. Egger, C.H. Mak, and U. Weiss, J. Chem. Phys. 100 (1994) 2651.

\bibitem{carmeli}
B. Carmeli and D. Chandler, J. Chem. Phys.
82 (1985) 3400; {\sl ibid.} 89 (1988) 452.


\bibitem{stock3}
G. Stock and M. Thoss, Phys. Rev. Lett. 78 (1997) 578.

\bibitem{miller}
H. Wang, X. Song, D. Chandler, and W.H. Miller, J. Chem. Phys. 110
 (1999) 4828.

\bibitem{golosov2}
A.A. Golosov, R.A. Friesner, and P. Pechukas,
J. Chem. Phys. 112 (2000) 2095.

\bibitem{thoss}
H. Wang, M. Thoss, and W.H. Miller, J. Chem. Phys. 115 (2001) 2979.

\bibitem{lucke3}
A. Lucke, C.H. Mak, and J.T. Stockburger, J. Chem. Phys. 111 (1999) 10843.


\bibitem{wang}
H. Wang, J. Chem. Phys. 113 (2000) 9948.

\bibitem{nancy}
N. Makri and D.E. Makarov, J. Chem. Phys.
 102 (1995) 4600.

\bibitem{winterstetter}
M. Winterstetter and W. Domcke, Chem. Phys. Lett.
 236 (1995) 445.

\bibitem{stockburger1}
J. Stockburger and C.H. Mak,
Phys. Rev. Lett. 80 (1998) 2657. 
\bibitem{stockburger2}
J. Stockburger and H. Grabert,
Phys. Rev. Lett. 88 (2002) 170407.

\bibitem{costi1}
T.A. Costi and C. Kieffer, Phys. Rev. Lett. 76 (1996) 1683.

\bibitem{schoeller}
M. Keil and H. Schoeller, Phys. Rev. B 63 (2001) 180302.

\bibitem{sudip}
S. Chakravarty and J. Rudnick, Phys. Rev. Lett. 75 (1995) 501.

\bibitem{voelker}
K. V\"olker, Phys. Rev. B 58 (1998) 1862.

\bibitem{bailey}
D. Bailey, M. Hurley, and H.K. McDowell,
J. Chem. Phys. 109 (1998) 8262.

\bibitem{acp}
C.H. Mak and R. Egger, Adv. Chem. Phys. 93 (1996) 29, and references
therein.

\bibitem{VCM}
G.A. Voth, D. Chandler, and W.H. Miller, J. Chem. Phys. 93
(1989) 7009.

\bibitem{hanggi}
P. H\"anggi, P. Talkner, and M. Borkovec, Rev. Mod. Phys. 62 (1990) 251.

\bibitem{chandbook}
D. Chandler, {\sl Introduction to Modern Statistical Mechanics}
(Oxford University Press, 1987).

\bibitem{miller_calcaterra}
J.R. Miller, L.T. Calcaterra, and G.L. Closs, 
J. Am. Chem. Soc. 106 (1984) 3047.

\end{thebibliography}
\end{document}